\documentclass[onecollarge,natbib]{svjour2}

\bibpunct{[}{]}{,}{n}{}{,} 
\smartqed  
\usepackage{graphicx}
\usepackage{subfig}

\usepackage{mathptmx}   
\usepackage{multirow}
\usepackage{amsmath}
\usepackage{amssymb}
\usepackage{comment}

\usepackage{lineno}

\newcommand{\beq}{\begin{eqnarray}}
\newcommand{\eeq}{\end{eqnarray}}
\newcommand{\be}{\begin{eqnarray*}}
\newcommand{\ee}{\end{eqnarray*}}

\def\lsim{\raise0.3ex\hbox{$<$\kern-0.75em\raise-1.1ex\hbox{$\sim$}}}
\def\gsim{\raise0.3ex\hbox{$>$\kern-0.75em\raise-1.1ex\hbox{$\sim$}}}

\def\beq     {\begin{equation}}
\def\eeq     {\end{equation}}

\def\beq     {\begin{equation}}
\def\eeq     {\end{equation}}

\usepackage{ifpdf}
\ifpdf
\usepackage[pdftex]{hyperref}
\else
\usepackage[hypertex]{hyperref}
\fi

\hypersetup{
  pdftitle={Prospects for quarkonium measurements  in p--A and A--A collisions at the LHC},%
  pdfauthor={J.P. Lansberg, S.J. Brodsky, F. Fleuret, C. Hadjidakis},%
  pdfsubject={},%
  pdfkeywords={Quarkonium Production, Fixed-Target experiment, Large Hadron Collider},%
  pdfstartview={},%
  bookmarksopen=true, breaklinks=true, debug=true, %
  colorlinks=true, linkcolor=red, citecolor=blue, urlcolor=blue
}

\journalname{Few Body Systems}
\title{Prospects for quarkonium measurements  in p--A and A--A collisions at the LHC}

\author{Michael Winn}

\institute{Physikalisches Institut, Universit\"at Heidelberg until May 2016
  \\
 Laboratoire de l'Acc\'el\'erateur Lin\'eaire at Orsay since June 2016} 

\begin{document}

\maketitle

\begin{abstract}
The potential of the ALICE, ATLAS, CMS and LHCb detectors for the measurement of quarkonium in heavy-ion collisions, both in nucleus-nucleus~(A--A) and in proton-nucleus~(p--A) interactions, in the years 2015 until about 2030 in the LHC Runs 2, 3 and 4  with larger statistics and detector upgrades is described. A selection of newly available observables is discussed.  
\end{abstract}
\vspace{1cm}

\section{Introduction}
Quarkonium has been identified as a sensitive probe of deconfinement in a seminal contribution by Matsui and Satz~\cite{Matsui:1986dk} to heavy-ion physics in 1986.  In the last thirty years, the field has progressed significantly on the theoretical and the experimental side. 
 Quarkonium observables are certainly one of the drivers of the heavy-ion physics programme at the LHC and a key sector  to be studied in the future with higher precision thanks to larger available statistics and better resolution by all four large experiments. The experimental results at the LHC from Run 1, i.e., the data takings in the years 2010-2013, contributed significantly to the most recent progresses in the field.  
First, the LHC highlights and the still open questions are summarised in order to provide the context for future opportunities.
A detailed review of the first data taking campaign until the beginning of 2013 and the theoretical context beyond the scope of this article can be found in Ref.~\cite{Andronic:2015wma}. 

\section{Run 1 results and first Run 2 glimpses}
At the LHC, the CMS collaboration provided the first measurements separating the individual $\Upsilon$ states in A--A collisions~\cite{Chatrchyan:2011pe,Chatrchyan:2012lxa,Khachatryan:2016xxp} and showing clear evidence for a suppression pattern where the higher mass states are strongly suppressed. However, the combination of the data from CMS~\cite{Chatrchyan:2012lxa} measuring in the rapidity range $|y|<2.4$ and ALICE~\cite{Abelev:2014nua} measuring in $2.5<y<4.0$  on the $\Upsilon$(1S) production, could not be explained yet in a satisfactory way based on models on the market, see for a discussion e.g. in Ref.~\cite{Andronic:2015wma}.   

In the charmonium sector,  ALICE presented a clear sign of a  low transverse momentum ($p_T$) J/$\psi$ production component leading to a larger nuclear modification factor $R_{AA}$~\cite{Abelev:2012rv,Abelev:2013ila,Adam:2015rba,Adam:2015isa,Adam:2016rdg}, i.e., a weaker suppression compared to the results at RHIC by PHENIX~\cite{Adler:2003rc} and STAR~\cite{Adamczyk:2013tvk,Adamczyk:2016srz} for the  $p_T$ integrated results and at low $p_T$. This increase has been widely interpreted as a signature of late-stage J/$\psi$ production either during the life-time of the deconfined medium produced in heavy-ion collisions proposed in Ref.~\cite{Thews:2000rj} or at the phase boundary between hadronic and Quark-Gluon Plasma matter~\cite{BraunMunzinger:2000px}. This production mechanism is seen as a signature of deconfinement\footnote{The idea of late-stage J/$\psi$ production was already brought up by Matsui~\cite{Matsuirecomb} in 1987. Back then, it was considered as  an unfortunate background for the expected suppression due to deconfinement.}. In addition, ATLAS~\cite{Aad:2010aa}, CMS~\cite{Chatrchyan:2012np,CMS:2012vxa,Khachatryan:2016ypw} and ALICE measured at larger $p_T$ stronger suppressions than ALICE at low $p_T$. The experimentally more challenging $\psi$(2S) plays a crucial role to establish the physics picture of the different models and providing also discriminative power between them, see for instance in~\cite{Abelevetal:2014cna}. First measurements of the excited $\psi$(2S) have been published currently strongly limited by statistics~\cite{Khachatryan:2014bva,Adam:2015isa,Sirunyan:2016znt} awaiting a coherent phenomenological interpretation. The degree of thermalisation of the quarkonium states in A--A collisions can be also probed by measuring their azimuthal correlation in momentum space with respect to the bulk of soft particles emitted in the collision.  This azimuthal distribution can be described by a Fourier decomposition. Most prominently, a positive second Fourier coefficient $v_2$ is expected in case of (partial) thermalisation at intermediate and low transverse momentum in semi-central collisions.  A hint of azimuthal anisotropy of inclusive J/$\psi$ has been observed by ALICE with Run 1 data~\cite{ALICE:2013xna} at forward rapidity ($2.5<y<4.0$). CMS could detect a finite $v_2$ for prompt J/$\psi$ at midrapidity ($|y|<2.4$) for $p_T>6.5$~GeV/$c$ and down to  $p_T>3$~GeV/$c$ for $1.6<|y|<2.4$~\cite{Khachatryan:2016ypw}.  In summary, the measurements point to a certain degree of thermalisation at low and intermediate $p_T\lesssim 5$~GeV/$c$, but they are at the moment not able to disentangle between models considering production and destruction of charmonia during the medium lifetime or the statistical hadronisation model assuming the production at the phase boundary with thermal weights.  The imprecise knowledge of the initial charm quark abundance is in this context the largest uncertainty independent of the model class.  Finally, an additional J/$\psi$ production component has been detected at very low transverse momentum, $p_T\lesssim 0.3$~GeV/$c$, presumably from photon-induced origin in analogy to the production in ultra-peripheral collisions~\cite{Adam:2015gba}.                                           

In the light of this panorama, the future key questions in A--A collisions at the LHC are the following: To which extent do charm quarks or maybe even beauty quarks thermalise and what is the best framework to describe it dynamically? Which ingredients are crucial for the interpretation of the experimental data? What is the fate of a $c\bar{c}(b\bar{b})$-quark pair that would become a bound state in a pp collision in the high temperature phase of a heavy-ion collision? Which fraction will still end up in a bound state? What can high-$p_T$ quarkonium tell us about energy loss?

In p--A collisions, there is not yet a consensus of the most important effects, which produce deviations from the naive nucleon-nucleon collision scaling expectation for quarkonium production at the LHC. The use of nuclear parton distribution functions (nPDF's) even available at NLO~\cite{ Hirai:2007sx, Eskola:2009uj,deFlorian:2011fp, Kovarik:2015cma}, in practice most calculations employ the EPS09 parameterisations~\cite{Eskola:2009uj},  can at least approximately account for the measured J/$\psi$~\cite{Aaij:2013zxa,Aad:2015ddl,Abelev:2013yxa,Adam:2015iga} and the $\Upsilon$(1S)~\cite{Aaij:2014mza,Abelev:2014oea} nuclear modification factors.
 Nevertheless, the currently available gluon distributions in the relevant region of Bjorken-$x$ suffer from large uncertainties already for the nucleon but even more for the nucleus in absence of lepton-ion collider data or other sensitive inputs. In addition,  it appears unclear whether the collinear factorisation inherent to the use of the ’standard’ nPDF's is appropriate to describe low $p_T$ charmonium~\footnote{Although bottomonium production has been also calculated within saturation approaches, the collinear framework should work more reliably in this case due to the larger probed Bjorken-$x$.} production at the LHC.

 Saturation due to large phase space occupancy of the partons might become important for the charm sector at low $p_T$, even more in collisions involving nuclei than in pp due to the $A^{1/3}$ dependence of the saturation scale~\cite{Fujii:2006ab}. Calculations in the Color Glass Condensate~(CGC) framework have been performed for the nuclear modification factor~\cite{Fujii:2013gxa,Ma:2015sia,Ducloue:2015xqa} at forward rapidities. Although the first attempt in Ref.~\cite{Fujii:2013gxa} failed to describe the data, recent calculations describe the experimental data reasonably well~\cite{Ma:2015sia,Ducloue:2015xqa}.
 
Alternatively, it was proposed that the main mechanism for the modification of charmonium and bottomonium could be coherent energy loss due to small angle gluon radiation of the colour dipole traversing the nucleus~\cite{Arleo:2012rs}. This model could also reproduce the main features of the available data for J/$\psi$~\cite{Abelev:2013yxa,Aaij:2013zxa,Adam:2015iga} and $\Upsilon$(1S)~\cite{Aaij:2014mza,Abelev:2014oea} production at low $p_T$, where the strongest modifications are observed both in data as well as in the model.  

In summary, a variety of models  exists and there is at the moment no conclusive and broadly accepted interpretation of the available data. Therefore, it appears not advisable to use the p--A data on quarkonium to constrain the nPDF's. It is also not possible to conclude that the coherent energy loss is the main effect or to claim that CGC calculations provide the best description of the data. Extrapolations to the expected non-deconfined nuclear modifications in A--A collisions based on p--A data 
can only be done under assumptions correct within certain frameworks.  Nevertheless, first data driven attempts were pursued~\cite{Adam:2015iga}, which prove the power of the precise measurements in p--A collisions to constrain the expectations in A--A collisions if the underlying assumptions are valid.    

Interestingly, the $\psi$(2S) is stronger suppressed than the J/$\psi$ in p--A collisions at the LHC~\cite{Abelev:2014zpa,Aaij:2016eyl}.  This behaviour lacks an explanation within the nuclear parton distribution framework and the coherent energy loss model. The comover model  which incorporates a late-stage interaction of the $c\bar{c}$ pair 
leading to the suppression of the excited state could as a first model account for the experimental observation~\cite{Ferreiro:2014bia}. The measured behaviour of $\psi$(2S) production in A--A collisions~\cite{Khachatryan:2014bva,Adam:2015isa} was also interpreted as a sign of late-stage interactions like in p--A collisions within  an in-spirit similar approach than the comovers and a later freeze-out of the $\psi$(2S)~\cite{Du:2015wha}.  
This interpretation will need to be reassessed with the Run 2 data, which starts to become available~\cite{Khachatryan:2014bva}.
In addition, a decrease of the yield ratio between the excited $\Upsilon$ states normalised to the ground state as a function of midrapidity multiplicity  has been observed  in p--A and in pp~\cite{Chatrchyan:2013nza} data. This observation might be related to the same physics as the $\psi$(2S) suppression. In addition to measurements as a function of multiplicity in pp~\cite{Abelev:2012rz,Chatrchyan:2013nza} as well as in p--Pb~\cite{Chatrchyan:2013nza,MartinBlanco:2014pia},  centrality-differential measurements in p--A collisions aiming at the distillation of the purely geometrically induced effects have been put forward in two publications by the ALICE Collaboration~\cite{Adam:2015jsa,Adam:2016ohd} and in preliminary results by ATLAS~\cite{ATLAS:2015pua} using different approaches. The ALICE measurement shows a centrality dependence of the stronger suppression of $\psi$(2S). The J/$\psi$ modification itself is roughly compatible with the mechanisms also describing the data as a function of rapidity and $p_T$. A centrality dependence of the $\psi(2S)$ to J/$\psi$ ratio is also visible in the ATLAS result at  $p_T>10$~GeV/$c$. A quantitative comparison of data and theory as a function of centrality in p--A collisions and a clear interpretation of the results will however require a further understanding of the centrality definition with its large intrinsic resolution and systematic uncertainties and its mapping in theoretical calculations in p--A collisions. 

In summary, the following questions arise from the p--A quarkonium programme at the LHC:  Do we need to treat charmonium production at low $p_T$ with a framework taking into account saturation physics? What mechanism does exactly cause the break-down of factorisation for the excited quarkonium states? Can we identify the degrees of freedom and the time scales of the responsible process? Can quarkonium help to understand the findings in the correlation sector in pp and p(d)--A collisions at high-multiplicity by providing a mass scale and therefore a infrared cut-off in the process? 

In the future, the experimental program  will profit from higher collision energies, larger delivered luminosities by the LHC accelerator complex and larger experimental efficiency due to detector and data acquisition upgrades as well as improvements in experimental resolutions. 
In particular, the number of accessible observables will grow and the achieved precision in already established measurements will strongly improve mainly thanks to larger heavy-quark statistics. After a general introduction about the quarkonium capabilities of the 4 large LHC experiments, explanations about the expected improvements by instrumental upgrades and the increase in statistical reach are given. Finally, potential new measurements and comparisons with the open heavy-flavour sector and potentially Drell-Yan will become feasible, which were not yet accessible during Run 1.

\section{Time schedule and expected delivered luminosities}
\begin{figure}[!htb]
  \begin{center}
\includegraphics[width= 0.5 \textwidth]{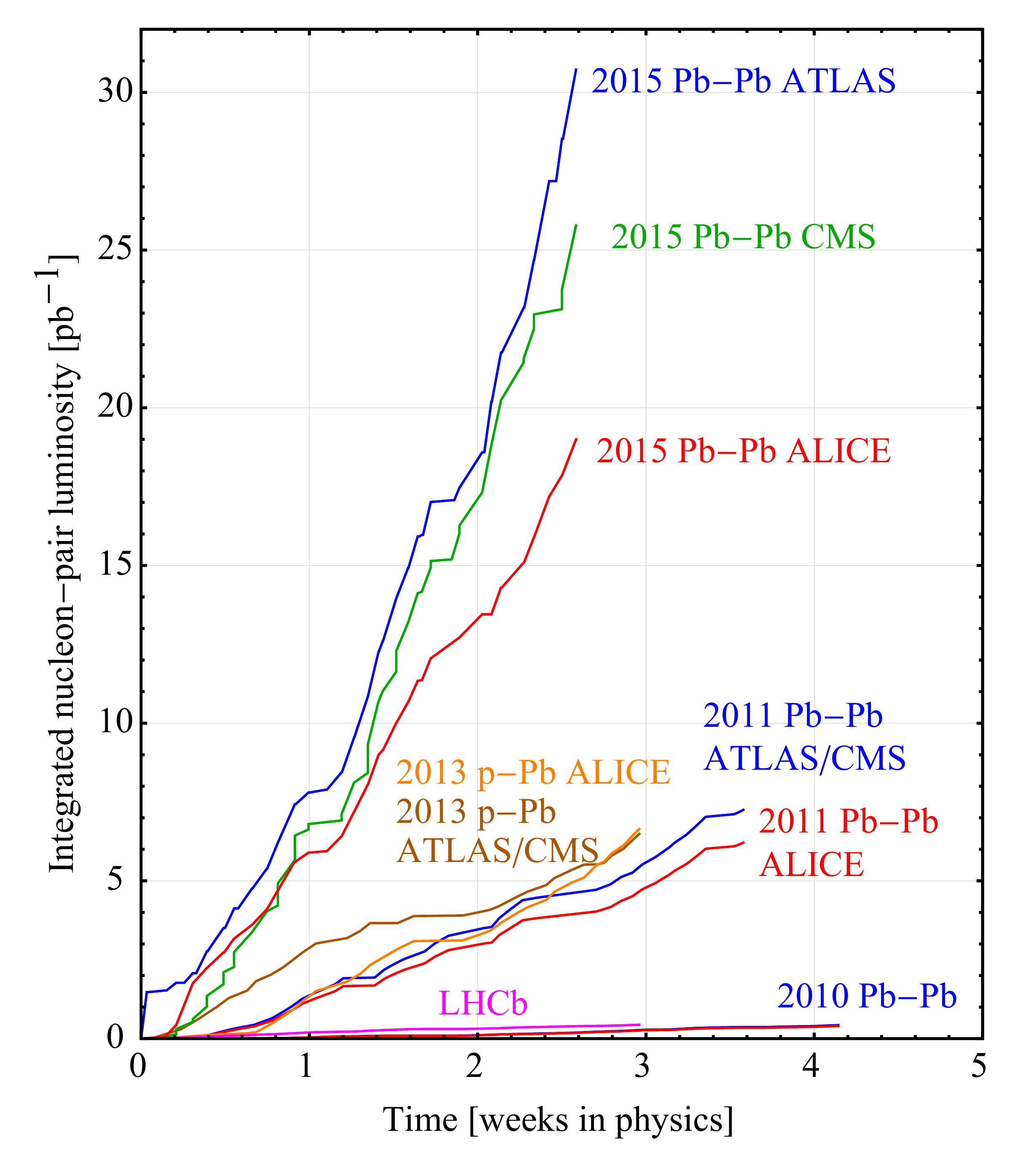}
\caption{LHC performance in terms of accumulated integrated equivalent nucleon-pair luminosity in different collision systems taken from Ref.~\cite{Jowett:2016ctg}. 
\label{fig:lumi}
}
\end{center}
\end{figure}

In Fig.~\ref{fig:lumi}, the delivered luminosities in the past heavy-ion data taking periods are summarised including the 2015 Pb--Pb data taking. The luminosities in the different systems are translated into equivalent luminosity for hard process observables in absence of nuclear effects, i.e., multiplying the delivered luminosity by $A=208$ in p--A collisions and by $A^2$ in A--A collisions. 
During the p--A data taking in 2016, a period with low interaction rate at $\sqrt{s_{NN}}=$~5 TeV is foreseen specifically for ALICE in order to collect 1 billion minimum bias collisions corresponding to about 500 $\mu b^{-1}$. Data samples of 100~nb$^{-1}$ are requested for CMS/ATLAS as well as 20 nb$^{-1}$ for LHCb at $\sqrt{s_{NN}}=$8 TeV. A beam direction inversion is foreseen to allow to test the proton and the lead fragmentation region by LHCb and the ALICE forward arm. A second Pb--Pb data taking period is foreseen at the end of Run 2 (2018), during which an integrated luminosity similar to the one delivered in 2015 can be expected.

\begin{figure}[!htb]
  \begin{center}
\includegraphics[width= 1.0 \textwidth]{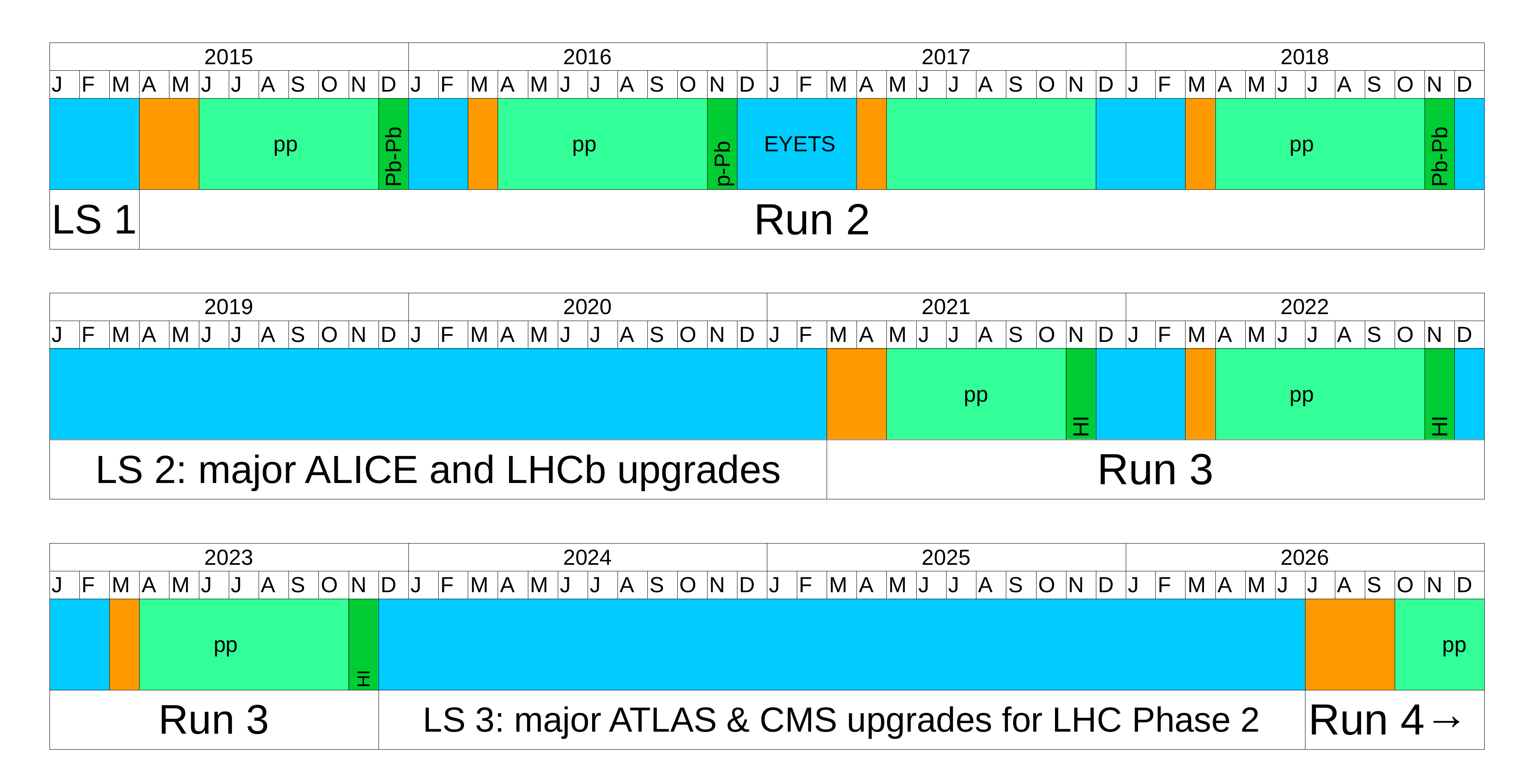}
\caption{The mid-term schedule for the LHC running. The information is taken from~\cite{timingwebpage}. Light blue indicates longer shut-down periods, orange commissioning phases and green actual physics production phases. Short technical stops are not indicated. The exact timing in particular after the end of Run 2 is subject to operational changes. 
\label{fig:schedule}
}
\end{center}
\end{figure}

 Based on the physics case outlined in the Letter of Intent of the ALICE upgrade~\cite{Abelevetal:2014cna}, ALICE requests the delivery of  10 nb$^{-1}$ Pb--Pb collisions at an interaction rate of 50 kHz  in Run 3 and Run 4. This request is the baseline scenario, which is also the goal for the improvements of the accelerator performance thanks to upgrades to be installed during the second long shut-down in 2019/2020. The 2015 accelerator performance exceeded the design luminosity and reached  already 35\% of the upgrade goal~\cite{Jowett:2016ctg}.  We assume in the following that the same or larger luminosities will be delivered to ATLAS and CMS compared to ALICE, since both detectors are able to cope with rates exceeding the maximum interaction rate considered for ALICE. The mid-term schedule of the LHC programme up to the start of Run 3 including these upgrades is summarised in Fig.~\ref{fig:schedule}. A comparison of the luminosities obtained in Pb--Pb in 2015 and the anticipated numbers for a one year performance during Run 3 is shown in Table~\ref{tab:example}. 
 Since LHCb only joined recently the A--A programme, it is not possible to be  quantitative about the certainly strong potential of LHCb in heavy-ion collisions for quarkonium studies and the delivered luminosities.

\begin{table}[htbp]
\centering
\begin{tabular}{|c|c|c|}
\hline
Experiment & Run 2 performance: 2015 Pb--Pb & Run 3 Pb--Pb performance for one heavy-ion year   \\
\hline
ALICE central barrel  & 150 mio  MB evts. recorded (0.02 nb$^{-1}$) &  $\approx$ 22.5 bio  MB recorded (3 nb$^{-1}$)*     \\
ALICE muon arm        &  0.225 nb$^{-1}$ analysed  & $\approx$ 3 nb$^{-1}$*             \\
\hline
CMS          & 0.464 nb$^{-1}$ analysed     &   	$\gtrsim$~3nb$^{-1}$**              \\
ATLAS        & 0.515 nb$^{-1}$ analysed   & $\gtrsim$~3nb$^{-1}$**       \\
\hline
LHCb &  50 mio MB  evts. recorded, 50-100\% centr. with tracking &  ***    \\
\hline
\end{tabular}
\caption{2015 data taking luminosities and anticipated one-year-performance during Run 3 (2021-2023). The ALICE, CMS and ATLAS luminosity numbers for the 2015 data taking are taken from~\cite{Adam:2016rdg,Khachatryan:2014bva,Sirunyan:2016znt}. Minimum bias is abbreviated by ’MB’.
  \\
  (*) Numbers assuming an increase in read-out rate by 150 from 'recorded' numbers for central barrel, i.e., assuming that the accelerator meets the goal of a levelled delivered rate of 50~kHz.
  \\
  (**) Assuming that LHC will deliver a levelled rate of 50~kHz or higher rates to ATLAS and CMS.
  \\
  (***) LHCb only participated in 2015 in Pb--Pb data taking with a low number of colliding bunches.  The LHCb luminosity will have to be assessed based on the filling scheme including consequences for the other interaction points and the beam optics.
}
\label{tab:example}
\end{table}

 \section{Access to quarkonium at the LHC in heavy-ion collisions}
 Quarkonium\footnote{We refer here always only to states below the open charm/beauty threshold.} production measurements in hadronic collisions usually rely on the sizeable partial decay width of the vector states ($\psi$) to dileptons. In addition, $p$-wave states are accessible by measurements of the type $\psi+\gamma$, where the $\gamma$ is rather soft.  The clear distinction of the different states typically requires the reconstruction of the photon as a conversion\footnote{LHCb measured for the first time the lowest mass state in the charmonium system $\eta_c$  in pp collisions by using the $p\bar{p}$ decay channel~\cite{Aaij:2014bga}. A measurement based on this channel or using other fully hadronic decay channels like $\Lambda-\bar{\Lambda}$ might be also conceivable with the ALICE central barrel and LHCb in Run 3 thanks to the good hadron particle identification and mass resolution also for hadronic tracks in p--A, but maybe also in A--A collisions. Since no dedicated simulation studies were presented so far, this observable class isn't discussed here.}. The p-wave states also contribute significantly to the inclusive production of the J/$\psi$, the $\Upsilon$(1S) and the $\Upsilon$(2S) by feed-down. Hence, their behaviour is crucial for the interpretation of the  vector state measurements.

\begin{figure}[!htb]
   \begin{center}
\begin{minipage}[t]{0.49\textwidth}
\includegraphics[width=1.0\textwidth]{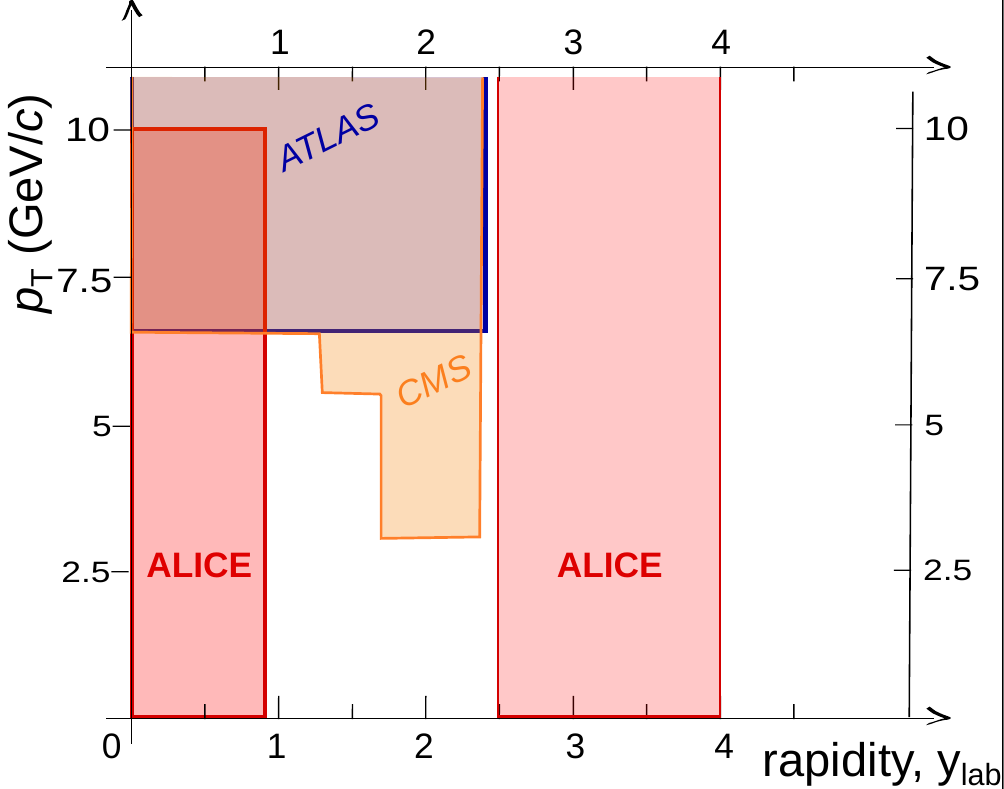}
\end{minipage}
\begin{minipage}[t]{0.49\textwidth}
\includegraphics[width=1.0\textwidth]{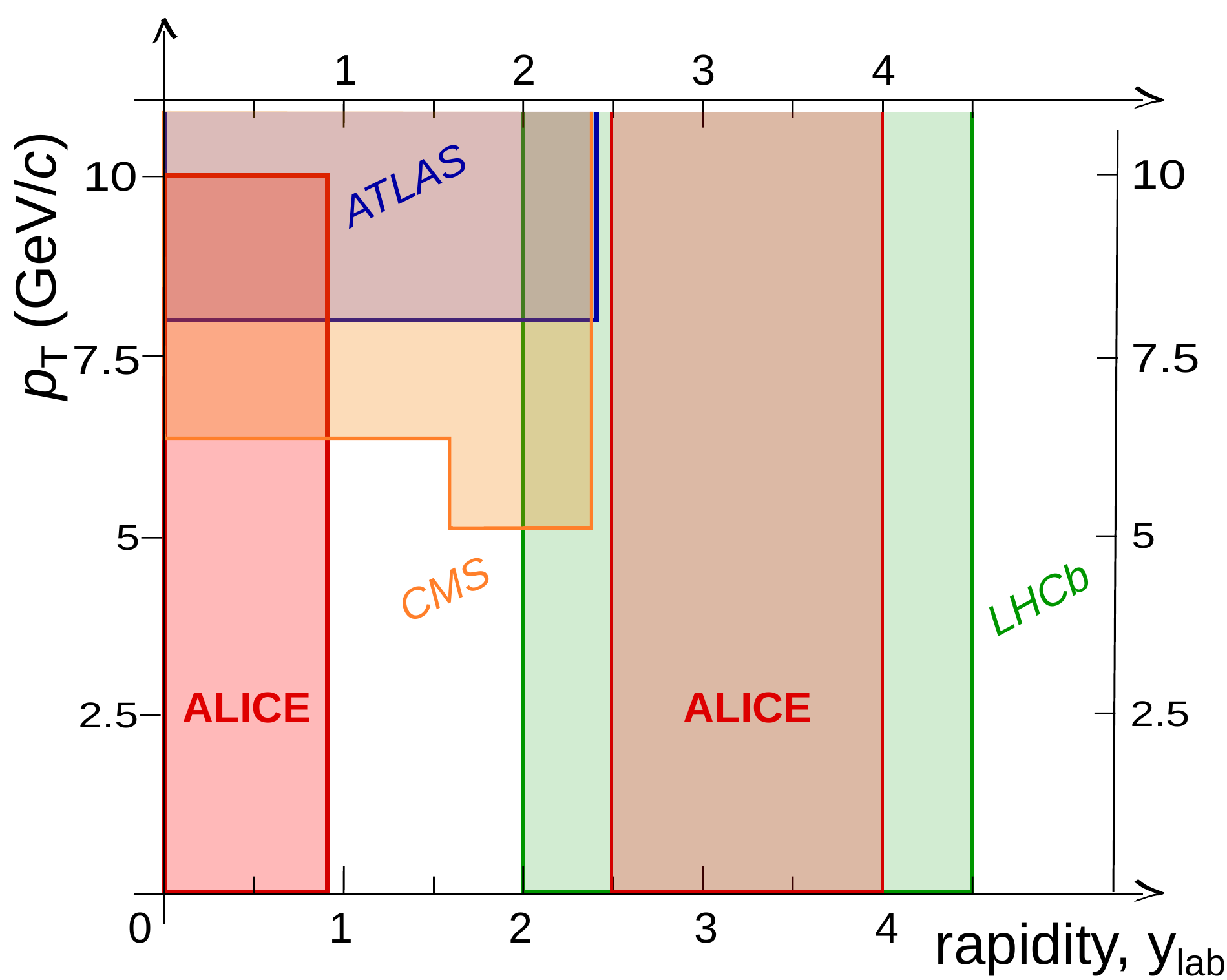}
\end{minipage}
\caption{Acceptance of J/$\psi$ measurements  by the four LHC experiments in Run 1 in Pb--Pb collisions (left) and p--Pb collisions (right). For the midrapidity detectors as the ALICE central barrel, ATLAS and CMS only one half of the acceptance is shown. The ALICE central barrel results from Run 1 are restricted to low transverse momentum due to statistical limitations.  References to the publications can be found in the text. 
  \label{fig:accjpsi}
}
\end{center}
\end{figure}

 In the upcoming data periods, all four experiments will take data in p--A as well as in A--A collisions. 
 In Fig.~\ref{fig:accjpsi}, the acceptance of the different detectors for the measurement of J/$\psi$  is shown as a function of $p_T$ and rapidity based on measurements in A--A or expected performance based on already available information and in the pp collision system as comparison in collider mode. For the $\Upsilon$ family, all experiments have acceptance down to zero $p_T$ of the dilepton pair within the same rapidity ranges. However, the resolution, the trigger and read-out capabilities of the detectors set practical limits.  In the following, we will discuss the potential of the different set-ups separately.

\subsection{ALICE muon arm}
The ALICE muon arm is placed at forward rapidity, $2.5<y<4.0$. It reconstructs the J/$\psi$, $\psi$(2S) and the $\Upsilon$ states in the dimuon decay channel down to zero $p_T$.
A large fraction of the delivered luminosity is available as recorded dimuon trigger sample. Therefore, it was already able to largely profit from the delivered luminosity by the LHC in Run 1.  The complete spectrometer  is placed behind a hadron absorber during Run 1 and Run 2 in order to keep the occupancy low in the muon stations even in most central collisions.
This design feature prevents from measuring the fraction of J/$\psi$ or $\psi$(2S) from B-hadrons by secondary vertexing, i.e., all measurements are always inclusive measurements. 

\begin{figure}[!htb]
  \begin{center}
\includegraphics[width= 0.5 \textwidth]{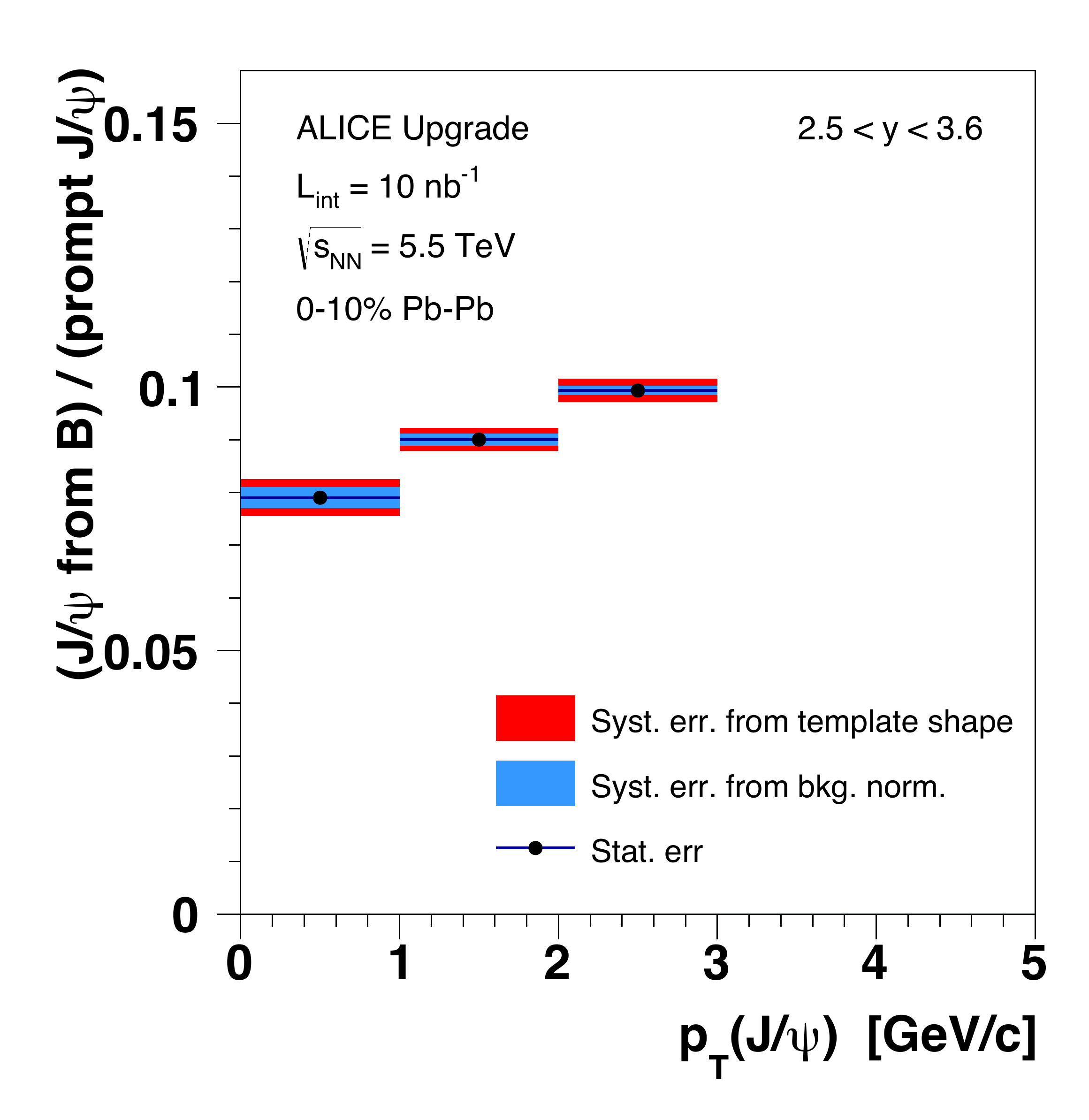}
\caption{Expected systematic uncertainties on the measurement of the displaced/prompt J/$\psi$ with the MFT of the ALICE muon arm taken from Ref.~\cite{CERN-LHCC-2015-001}. 
\label{fig:MFT}
}
\end{center}
\end{figure}

In 2019, in the second long shut-down, an upgrade is planned to supplement the spectrometer with silicon detector planes in front of the absorber~\cite{CERN-LHCC-2015-001}. This silicon detector, the Muon Forward Tracker~(MFT), will allow the separation of J/$\psi$ from b-decays in part of the acceptance ($2.5<y<3.6$) in conjunction with the primary vertex information from the central barrel detectors of ALICE. Fig.~\ref{fig:MFT}  shows the projection for the measurement of the fraction of the ratio between J/$\psi$ from B decays and prompt J/$\psi$.  In addition, the upgraded muon arm will see about a factor 10 more delivered luminosity (1 $nb^{-1}$ in Run 2 vs. 10 $nb^{-1}$ in Run 3 and 4) allowing for a precise measurement of J/$\psi$ $v_2$, an extension of the $p_T$ range for J/$\psi$ studies and studies of the excited $\Upsilon$ states despite the mass resolution limitations of the set-up.

\subsection{ALICE central barrel}
The ALICE central barrel measures J/$\psi$ in the dielectron channel in the rapidity range $|y|<0.9$.  The particle identification capabilities with the ALICE TPC and the ALICE TRD allow to measure J/$\psi$ down to zero $p_T$ at midrapidity which is unique at the LHC. These measurements are based on minimum bias or centrality triggers and are challenging due to the small signal-over-background ratios.  The main limitation of the set-up is the read-out rate limitation of about 200 Hz, when triggering primarily on the 0-10\% most central Pb--Pb events in 2011~\cite{Kolleger}, and about 1 kHz in p--Pb during Run 1 for low ’pile-up’ within the drift time of the TPC of about 90 $\mu s$. Due to this read-out rate limitation, the analyses are statistically limited and, e. g., the $\psi$(2S) could not be detected during Run 1. In the winter break between 2015 and 2016, the TPC read-out was upgraded in order to reach a 2 times faster read-out in Pb--Pb collisions~\cite{RCU2}. In p--Pb collisions, a read-out rate of about 1.6 kHz during the 2016 data taking period was achieved. 

The main upgrade of the ALICE central barrel is planned for the second long shut-down, in the period 2019-2020. One of the main pillars of the physics programme is the measurement of heavy-flavour including quarkonium down to zero $p_T$. The current 6 layer silicon tracker will be replaced by a 7 layer pixel-only layout. The detector will feature larger granularity, low material budget of $0.3 \%$ per layer for the innermost layers. The innermost layer is closer to the beam-pipe at $r= 22$~mm with a pixel size of O($30 \times 30$ $\mu$m$^2$) compared to in the current design of $50 \times 470~\mu$m$^2$ for the first two layers~\cite{Kim:2016ktw}. The read-out planes of the Time Projection Chamber~(TPC), at the moment equipped with multi-wire proportional chambers operated in gated mode to keep the ion back-flow and hence the space charge effects low, will be equipped with 4-stack GEM chambers. This set-up will have an intrinsic ion back-flow of 1~\% without gating and a similar d$E/$d$x$ performance than the present TPC~\cite{ALICE:2014qrd}. 
After these two major hardware upgrades in conjunction with an upgrade of the data acquisition and processing including the integration of partial online calibration, the detector is planned to be operated in continuous read-out in order to record all Pb--Pb collisions delivered by the LHC at a rate of 50 kHz. 
In p--Pb collisions, the future use of the  2015 completed transition radiation detector might enable a statistics increase of the ALICE programme at midrapidity down to lower $p_T$ than the ATLAS and CMS programmes.

\begin{figure}[!htb]
  \begin{center}
\begin{minipage}[t]{0.49\textwidth}
\includegraphics[width=1.0\textwidth]{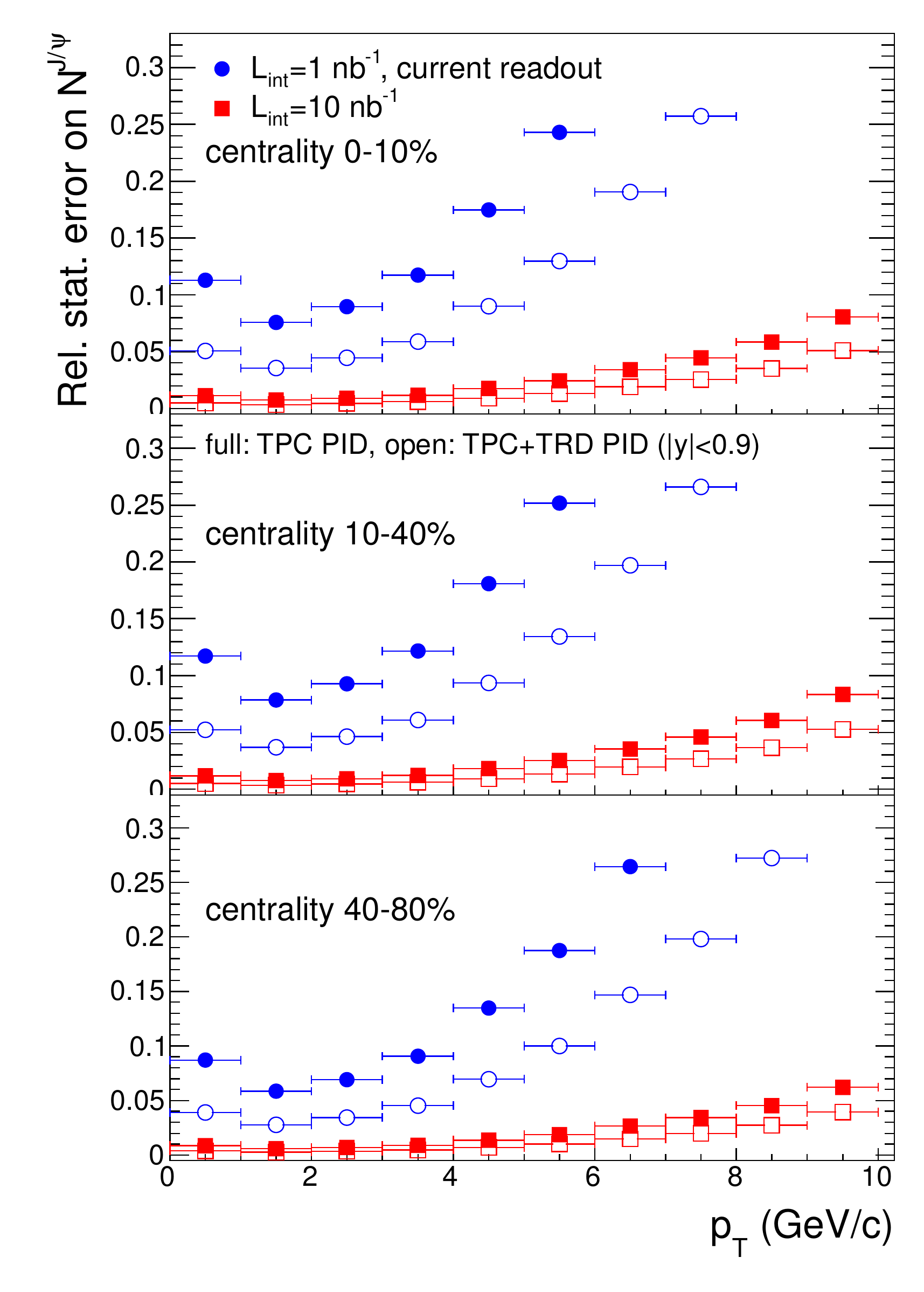}
\end{minipage}
\begin{minipage}[t]{0.49\textwidth}
\includegraphics[width=1.0\textwidth]{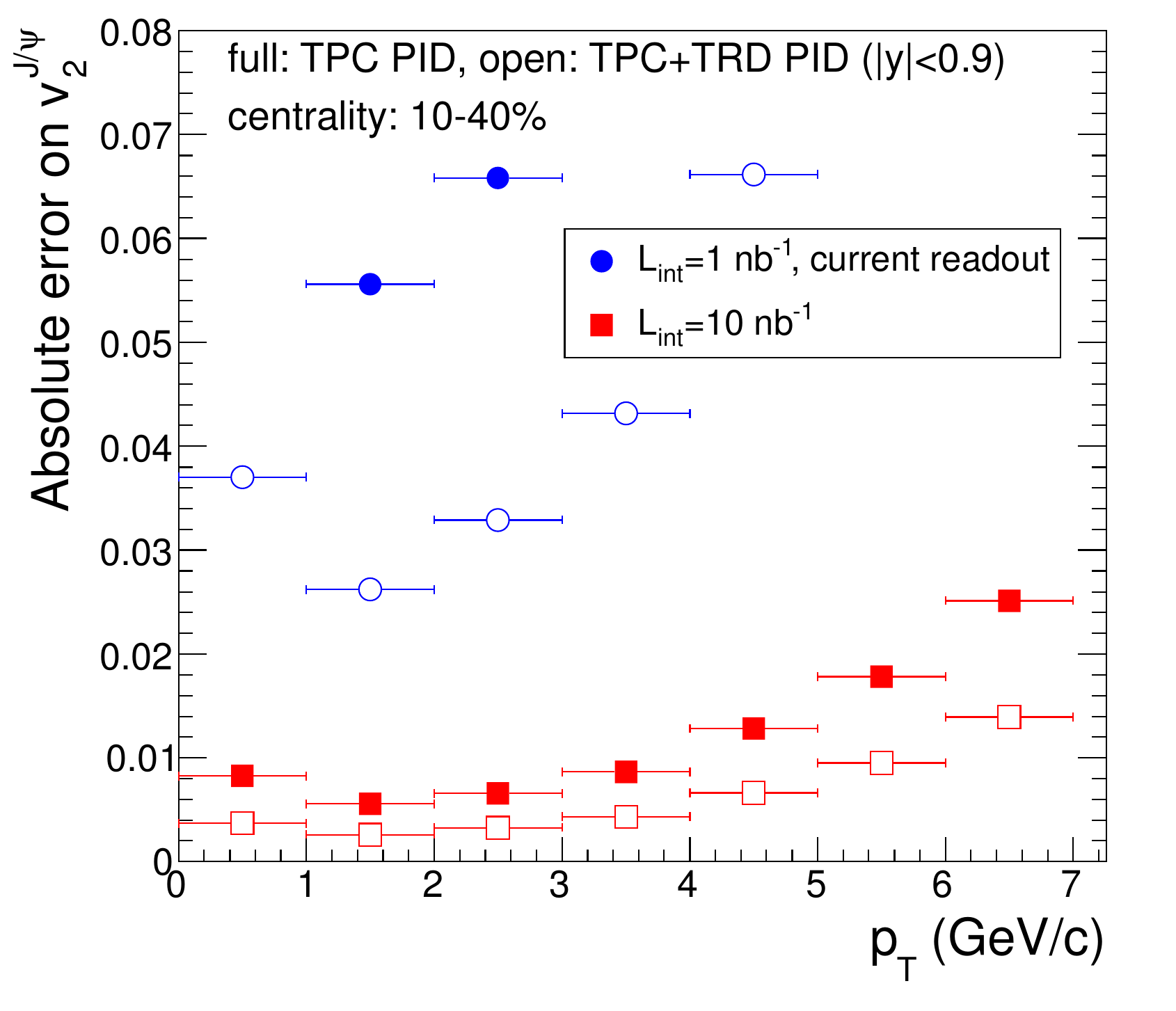}
\end{minipage}
\caption{Statistical uncertainties for the $p_T$ differential yield of J/$\psi$ (right hand side) in different centrality bins and for the elliptic flow of the J/$\psi$ in one centrality class with the ALICE central barrel with and without upgrade and for two different particle identification strategies in Pb--Pb collisions. The figure is taken from Ref.~\cite{Abelevetal:2014cna}.
  \label{fig:jpsicentbarrel}
}
\end{center}
\end{figure}

The ALICE central barrel upgrade will reduce the statistical uncertainties, since it will increase the available statistics compared to Run 2 by a factor O$(100)$. It  will allow for precision measurements of the nuclear modification factor of the J/$\psi$ as well as a measurement of the $\psi(2S)$ at midrapidity down to vanishing $p_T$.  A measurement of the elliptic flow of the J/$\psi$ down to low signal strengths will become possible. At the same time, the further reduction of the material budget will shrink the radiative tail in the dielectron channel due to Bremsstrahlung and improve the secondary vertex capabilities also thanks to the higher granularity. 
Figure~\ref{fig:jpsicentbarrel} shows a selection of the potential in terms of statistical uncertainties.

\subsection{Multi-purpose detectors}
At the LHC, the ATLAS and CMS detectors running at the maximal deliverable luminosities feature highly efficient muon-trigger systems with large acceptances, which are so far the basis for all quarkonium measurements presented in heavy-ion collisions.  In addition, the extended silicon vertex trackers allow for the separation of the prompt and the non-prompt component from B-hadron decays of the vector charmonium decays.  

\begin{figure}[!htb]
  \begin{center}
\includegraphics[width=.8\textwidth]{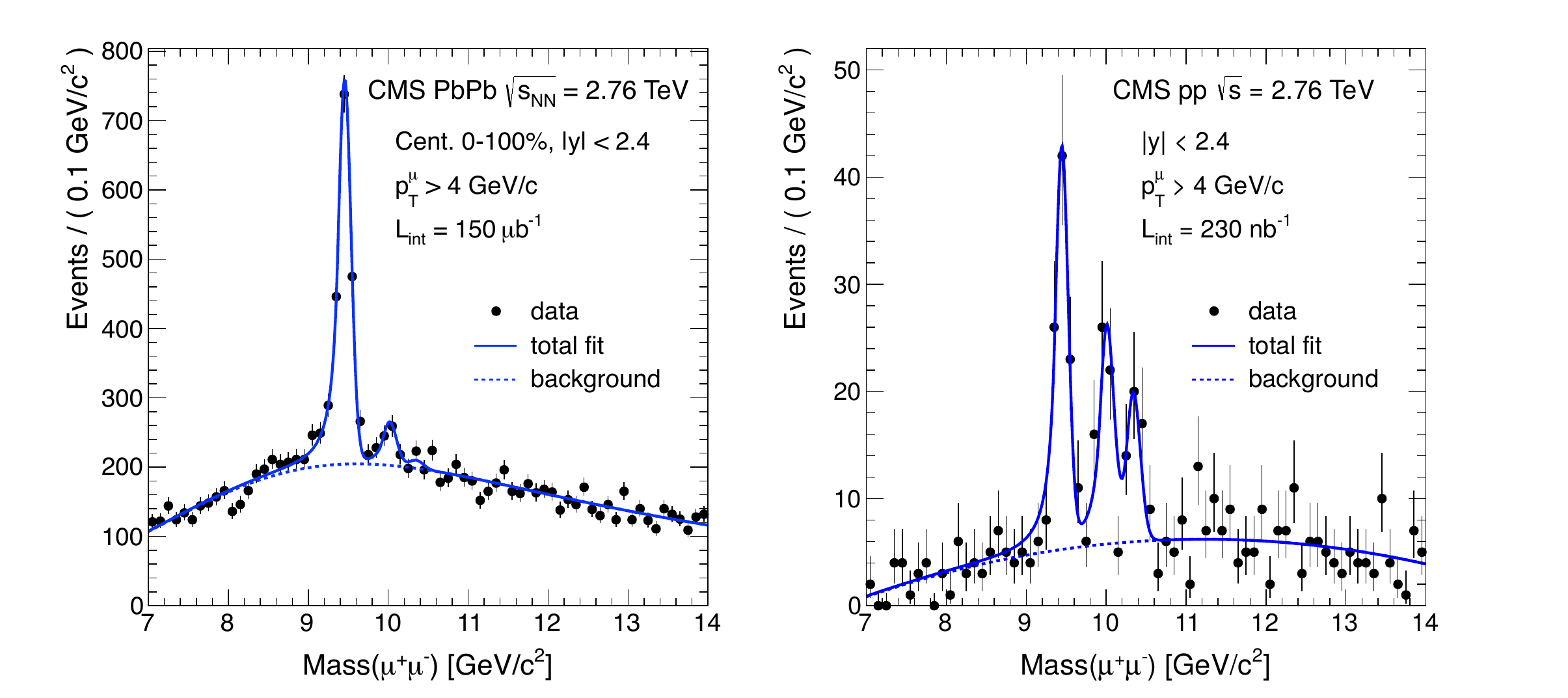}
\caption{Invariant mass distribution in Pb--Pb collisions and in pp collisions measured by the CMS collaboration during Run 1.  The Figure is taken from Ref.~\cite{Chatrchyan:2012lxa}.}
\label{fig:CMSUpsilon}
\end{center}
\end{figure}

The running at nominal magnetic field implies a single track cut-off at around 3.5-4.0 GeV/$c$ on the single muon seen in the muon systems at midrapidity. For the bottomonium system, this restriction allows to reach zero $p_T$ for the dilepton pair with large acceptance  factors. For the charmonium, the acceptance limit  implies a $p_T$ cut-off of about 7-10 GeV/$c$ for the dimuon $p_T$ in the measurements at midrapidity. At the forward rapidity edge of the muon systems, the acceptance of the dimuon trigger reaches lower in $p_T$ as demonstrated, e.g., in~\cite{Khachatryan:2014bva} also in Pb--Pb collisions.  The performance of the muon offline reconstruction even in most central Pb--Pb collisions allows for not strongly mitigated background conditions and similar resolutions as in $pp$ collisions for tight track selections in case of CMS. An example showing the performance for the $\Upsilon$ family measured by CMS in pp and Pb--Pb collisions during Run 1 is given in Fig.~\ref{fig:CMSUpsilon}.

\begin{figure}[!htb]
  \begin{center}
\includegraphics[width=.8\textwidth]{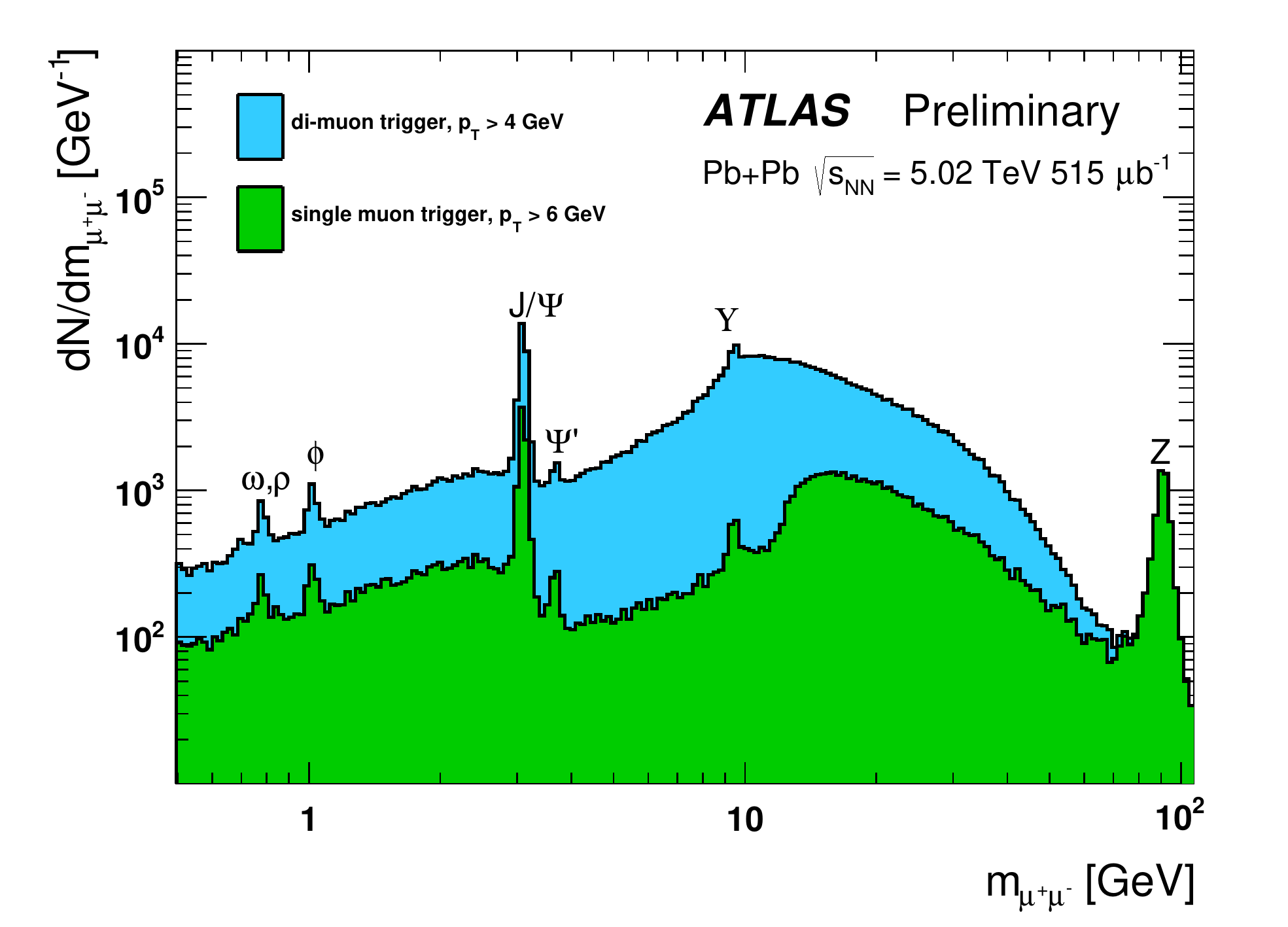}
\caption{Invariant mass distributions from the dimuon and single trigger lines by the ATLAS collaboration from the 2015 Pb--Pb data taking. The quarkonium vector state peaks are clearly visible~\cite{ATLASpub}.}
\label{fig:ATLASmass}
\end{center}
\end{figure}

Both detectors will largely profit from the increase of the heavy-ion luminosity for the dimuon trigger lines for quarkonium down to the low-$p_T$ end of the acceptance in Run 2 and Run 3.  A first glimpse of the statistical power of the results to come can be deduced from the dimuon invariant mass distributions from the 2015 \mbox{Pb--Pb} data taking campaign of the ATLAS collaboration shown in Fig.~\ref{fig:ATLASmass} and the first results already mentioned in the introduction. The exact increase of the available statistics will depend on the trigger menu set-ups. In Run 4, with the upgraded High Level Triggers (HLT's), the detectors will be able to cope with an input rate of 750 kHz from the L1 triggers for CMS~\cite{Contardo:2020886} and about 200 kHz for ATLAS~\cite{atlasup} in pp collisions. 
These upgrades will open the possibility to process a larger fraction the Pb--Pb interaction rate and a larger fraction of the event reconstruction in the HLT. 
A physics reach study for the projected 10 nb$^{-1}$ was performed by CMS~\cite{CMS:2013gga}. 
The number of produced $\Upsilon$(3S) in this data sample is comparable to the number of $\Upsilon$(1S) particles presently available from Run 1. 5500 raw prompt J/$\psi$ above $p_T>30$~GeV/$c$ integrated over centrality are estimated.  The measurement of the nuclear modification factor of the $\Upsilon$ family will be possible in steps of 10~\% centrality percentiles up to the most peripheral class in order to match smoothly the measurements in p--A and pp collisions. This measurement will allow to map out potential differences or similarities at the same track multiplicity in the different systems. Future measurements will certainly also profit from improved resolutions of the upgraded detectors as demonstrated for the $B_s \to \mu^+ \mu^-$ projections for CMS~\cite{CMS-PAS-FTR-13-022}. As already demonstrated with a first statistically limited study with 2011 data with prompt J/$\psi$~\cite{Khachatryan:2016ypw}, various azimuthal anisotropy measurements will  become feasible already with the new data collected in 2015.

 CMS and ATLAS proved for the J/$\psi$ to have acceptance coverage down to $p_{T}=0$~(CMS) or 1~(ATLAS)~GeV/$c$ in pp~collisions at a rapidity around $y=2$~\cite{Khachatryan:2010yr,Aad:2011sp}. In addition, the CMS collaboration demonstrated the power of their recorded number of minimum bias events in Pb--Pb collisions with D-mesons in the 2011 and 2015 data taking~\cite{CMS:2015hca,CMS:2016nrh} amounting to about 150 million events in 2015.  Certainly, minimum bias samples collected by the multi-purpose detectors could potentially allow measurements of charmonium extending  to lower $p_T$ than presently available data based on muons fully reconstructed in the muon systems. The exact reach will depend on the size of the achievable signal over background ratio and the trigger menu.

\subsection{LHCb}
\begin{figure}[!htb]
   \begin{center}
\begin{minipage}[t]{0.49\textwidth}
\includegraphics[width=1.0\textwidth]{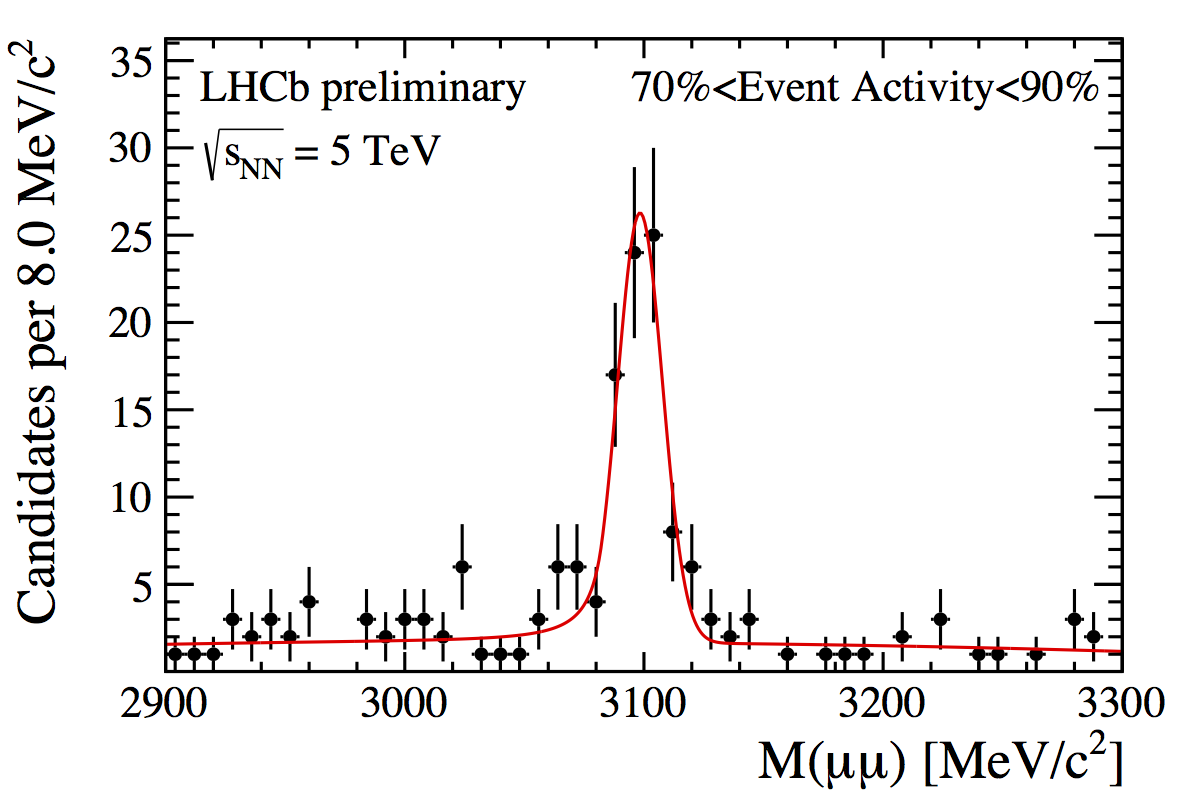}
\end{minipage}
\begin{minipage}[t]{0.49\textwidth}
\includegraphics[width=1.0\textwidth]{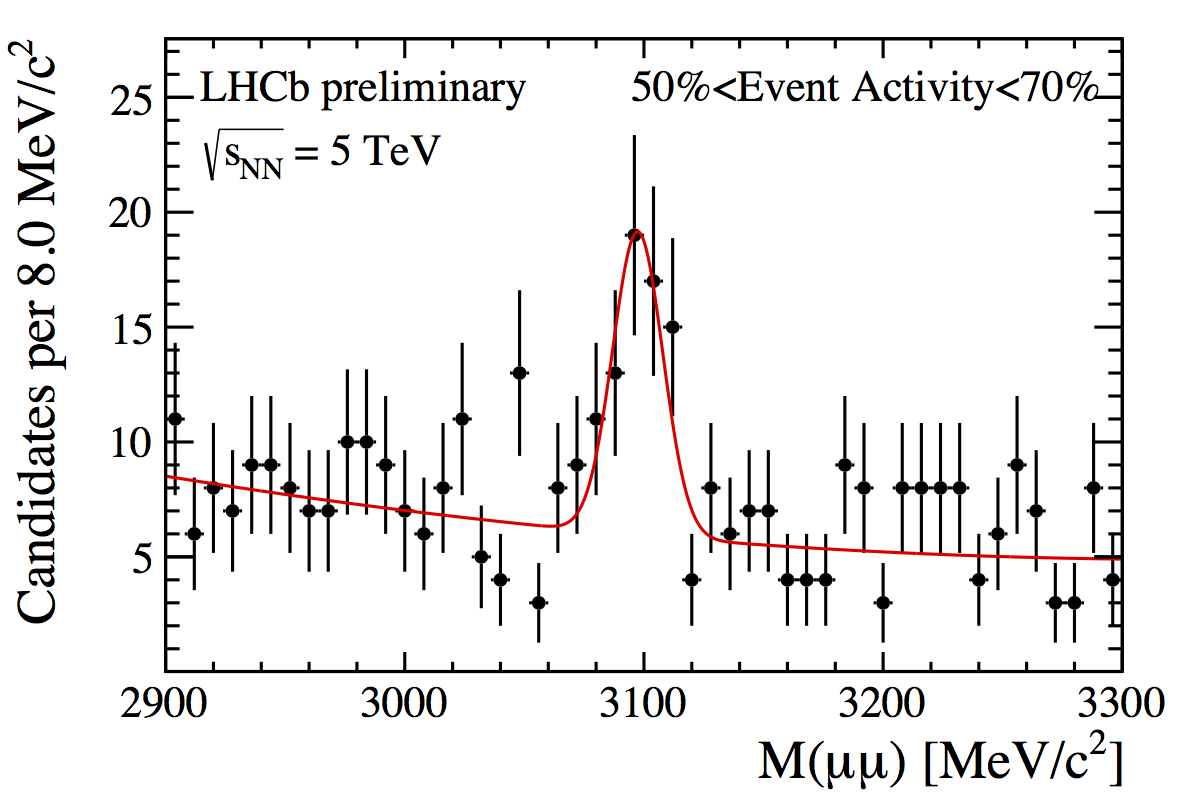}
\end{minipage}
\caption{Performance figures of the J/$\psi$ peaks seen by the LHCb collaboration in the 2015 Pb--Pb collision data sample taken from Ref.~\cite{LHCbwebPbPb}.
  \label{LHCb:pub}
}
\end{center}
\end{figure}

The LHCb detector took data for the first time in 2013 in p--Pb collisions and contributed already significantly to the measurement of non-deconfinement effects in nuclear collisions in the quarkonium sector~\cite{Aaij:2013zxa,Aaij:2014mza,Aaij:2016eyl}. The detector is fast, features a HLT with an input rate of 1 MHz in pp collisions and a large event rate storage capability of about 10~kHz. In Pb--Pb collisions, the lower tracker granularity does not allow to measure most central collisions. Currently, the tracking optimised for pp collisions allows to measure up to 50\% in centrality. A first idea of the performance in Run 2 based on the about 50 minimum bias collisions collected in 2015 can be deduced from the J/$\psi$~peaks shown in Fig.~\ref{LHCb:pub}. Measurements of $\chi_c$ states should be feasible in peripheral collisions based on the experience in pp collisions~\cite{LHCb:2012ac,LHCb:2012af}. LHCb allows within the event classes accessible to measure the full zoo of open heavy-flavour hadrons down to low $p_T$ thanks to the boost along the beam axis. The open charm species accesible to LHCb include the $\Lambda_c$. It was measured down to $p_T =$~2~GeV/$c$ in pp collisions~\cite{Aaij:2013mga}. This measurement represents the only published charm baryon cross section at the LHC to date.

The silicon vertex locator as well as the other tracker components will be completely replaced in the long shut-down of 2019-2020. The granularity of the detector will be strongly increased, since the detector upgrade set-up is foreseen to operate the detector at larger pile-up in pp collisions: instead of one visible collision  on average per bunch crossing, the design is planned to cope with 5.2 visible interactions to reach operations at an instantaneous luminosity of $2 \cdot 10^{33}$ cm$^{-2}$s$^{-1}$~\cite{Bediaga:1443882} in pp collisions. Furthermore, all bunch crossings will be processed by the HLT of LHCb.  This upgrade will be hence very beneficial for the centrality reach of the detector in Pb--Pb collisions. Detailed simulation studies and tracking optimisations need to be performed to estimate the reach in centrality. 

\subsection{LHCb as fixed-target experiment}
The LHCb detector has been equipped with a gas injection system to improve the luminosity determination~\cite{Aaij:2014ida}. The luminosity measurement is performed by imaging the beam by reconstructing the vertices originating from collisions of the beam particles with the injected noble gas. This method is complementary to the conventional Van Der Meer scans. The gas injection system increases the pressure in the LHCb interaction region vacuum by about 2  orders of magnitude by gas injections to about $O(10^{-7})$~mbar.

The system can be also used to study fixed target events between the proton or lead beam and the injected gas. LHCb becomes in this case a midrapidity detector slightly below the RHIC top collision energy. This option has been already exploited in 2015. Table~\ref{tab:fixtarg} summarises the data takings already performed with the corresponding energies. In particular for charmonium and more generally charm physics, the detector set-up offers a unique opportunity to provide measurements of nuclear effects in p--A collisions as well as in Pb--A collisions.  As an ultimate goal, it has been proposed to measure $\chi_C$ in Pb--Ar collisions up to most central collisions.  

\begin{table}[htbp]
\centering
\begin{tabular}{|c|c|c|c|}
\hline
Collision system & $\sqrt{s_{NN}}$ & data taking duration & rapidity coverage centre-of-mass frame     \\
\hline
p--A   &   &       &                                \\
\hline
p--Ne  &  86.6~GeV     & $\approx 30$~min   &  -2.5$<y_{cms}<$0.0 \\
p--Ne  &  110.4~GeV & $\approx 12$~h &  -2.8$<y_{cms}<$-0.3   \\
p--Ar & 110.4~GeV & $\approx 3$~days & -2.8$<y_{cms}<$-0.3 \\
p--Ar  &  69~GeV  & $\approx $ few~hours & -2.3$<y_{cms}<$0.2 \\
\hline
Pb--A &           &                 &               \\
\hline
Pb--Ne       &   55~GeV    & $\approx 30$~min &  -2.1$<y_{cms}<$0.4    \\
Pb--Ar &   69~GeV   &  $\approx 1.5$~weeks &   -2.3$<y_{cms}<$0.2  \\
\hline
\end{tabular}
\caption{Fixed target data samples relevant for heavy-ion physics taken by LHCb~\cite{LHCb:priv}.}
\label{tab:fixtarg}
\end{table}

\subsection{Non-quarkonium observables and their impact on quarkonium and new related observables}
Firstly, it is important to note that influences of potential (partial) thermalisation will affect the quarkonium production for  $p_T \leq O(m_{Hadron})$~\cite{Borghini:2005kd} where thermal equilibrium properties for heavy quark pair extracted from the lattice for heavy quarkonium will be most directly relatable to the experimental observables.  In this regime, heavy-quark observables are unique as the most direct access to quarks as colour-charge carriers implanted early on into the system without being destroyed or generated during the evolution.

This $p_T$ regime is at the same time the most challenging, but also a very interesting one for perturbative QCD calculations compared to pp and p--Pb data due to the probed low Bjorken-$x$ and the comparatively small size of the hard scale in the hard matrix element of a few GeV.   
In this context, the presence of a deconfined medium  is not expected to modify strongly the production of the heavy-quark states themselves, but only their density distribution in rapidity, $p_T$ or relative to the rest of particle production or in species distribution, i.e., the abundance of open, hidden heavy flavour mesons or baryons. The total $c\bar{c}$ and $b\bar{b}$ production rate is hence assumed to be independent of deconfinement effects. Measurements in p--A collisions are due to this reason the testing ground of all non-deconfinement effects present in A--A collisions. However, depending on the dominating physical effects,  also the production modification of total heavy-quark production itself and even more the production of more specific heavy quark observables might not factorise going from p--A to A--A easily. 

For the latter statement, the CGC framework is considered as an example. The dilute-dense approximation used for forward and partially also for midrapidity heavy quark production breaks down at midrapidity in heavy-ion collisions. It has to be replaced by  a ’dense-dense’ calculation, which cannot be  factorised in the elements appearing in the ’dilute-dense’ calculation. The calculation in the ’dense-dense’ framework was not yet done for heavy-quark production.

In this context,  the models for quarkonium production or at least their implications have to be exposed to the measurements of leptons from heavy-quark hadron decays and  fully  reconstructed open charm and beauty  hadrons accessible by ALICE and LHCb down to low or vanishing $p_T$ and by CMS and ATLAS up to very high-$p_T$. In addition, it is important that always all input assumptions  yielding to results and their interpretation on the phenomenology as well as on the experimental side are clearly stated and discussed.

\begin{figure}[!htb]
  \begin{center}
\includegraphics[width=.8\textwidth]{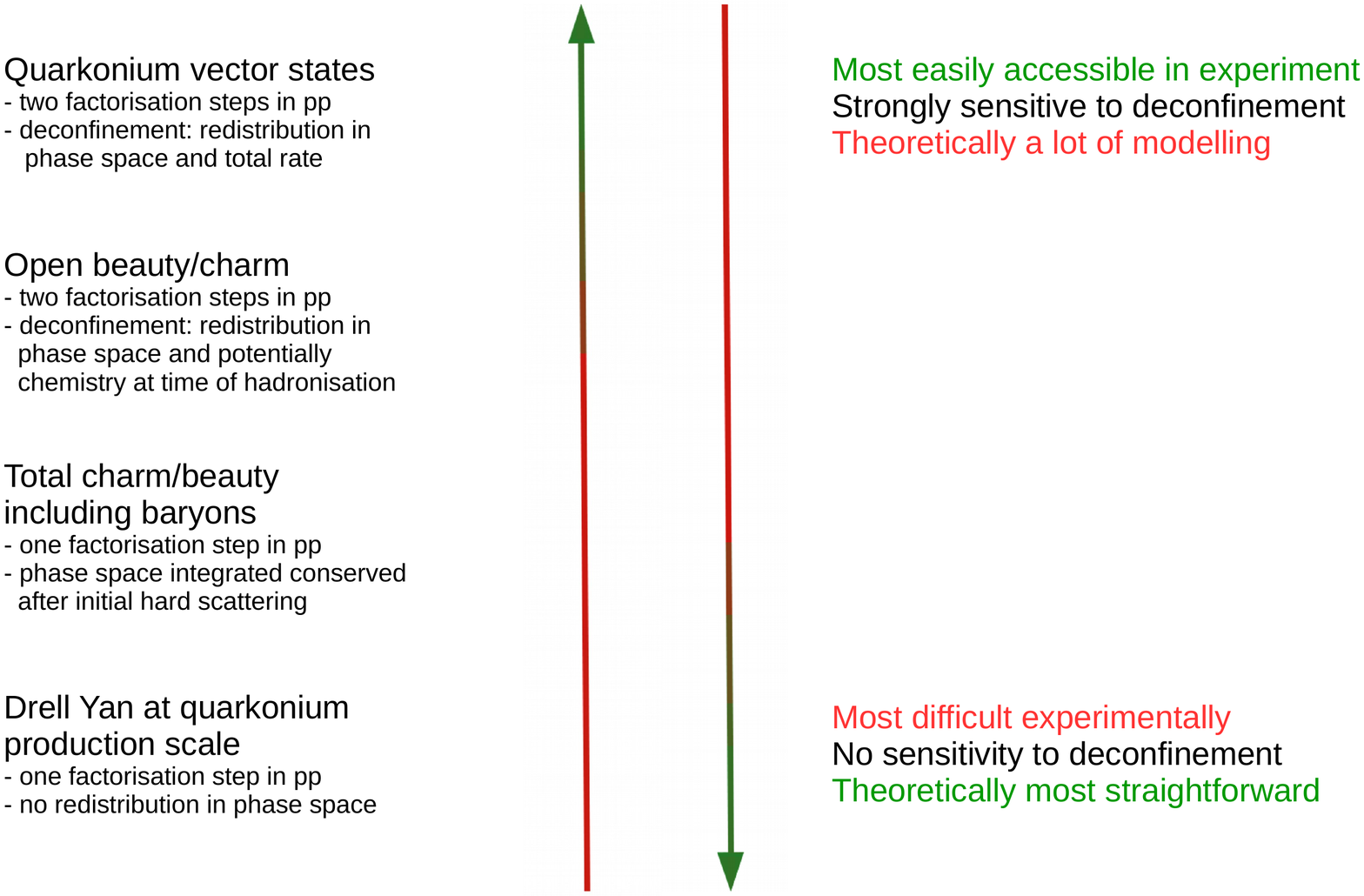}
\caption{Cartoon illustrating opportunities and challenges related to different observables.}
\label{fig:flow}
\end{center}
\end{figure}

 The measurement of ratios of different observables cancelling uncertainties in theory and/or in experiment, or at least their comparison, have the potential for new insights. However, it is clear that the observable itself might make only sense in a certain theoretical framework or might require more modelling work on the phenomenological side than bare yields, nuclear modification factors or cross sections, as illustrated in Fig.~\ref{fig:flow}. 
 Nevertheless,  a clear falsification or a confirmation of the expectation can provide large progress.  A few ideas are listed in this context:
\begin{enumerate}
\item Drell-Yan production between the $\psi$(2S) and the $\Upsilon$(1S) and the  J/$\psi$: this measurement was proposed in~\cite{Arleo:2015qiv} for p-A collisions. With the requested luminosity by LHCb in the 2016 running, this observable should be in reach in p--A collisions~\cite{Gracziani:2145943}. Certainly, there are experimental uncertainty cancellations. The cancellation  of scale variation uncertainties on the theory side was however controversial at this workshop. It would be also very interesting in AA collisions; the measurement would be even more challenging. 
\item J/$\psi$ over total open charm integrated over $p_T$  as a function of centrality in A--A and compared to p--A and pp\footnote{In absence of any rapidity shifts.}:  in a pure suppression picture, the production ratio integrated over $p_T$ is the golden observable to quantify the modification of J/$\psi$ and other quarkonium states. It has been also proposed in the context of late stage J/$\psi$ production models, e.g. in Ref.~\cite{Andronic:2006ky}. Experimentally, the measurement of $\Lambda_{c}$, which contributes to about 10\% of the total $c\bar{c}$ cross section in pp collisions, is certainly the main challenge and experimental uncertainty cancellations are limited. The measurement of $\Lambda_c$ down to low $p_T$ is one of the key measurements of the ALICE upgrade. 
\item J/$\psi$ over D-meson production as a function of centrality in A--A and compared to p--A and pp:  this measurement with the $p_T$ integrated quantities is certainly feasible with LHCb in peripheral collisions with the presently available data and might be also feasible with the present data by ALICE with extrapolation down to zero $p_T$ at midrapidity. The upgrade  of ALICE should allow to access the observable at midrapidity in A--A. There is certainly room for the cancellation of uncertainties in this ratio from an experimental point of view. Theoretically, the full exploitation of this ratio will require assumptions on the hadronisation of charm quarks.
\item $\Upsilon$~production over non-prompt J/$\psi$ as a function of centrality in A--A and compared to p-A and pp: it should allow to cancel uncertainties in the dimuon trigger systems of CMS, ATLAS or LHCb and on muon tracking in general.   The caveats in extrapolating to the $p_T$ integrated cross section  here are the extrapolation down to 0 $p_T$ for the non-prompt J/$\psi$ and the  assumptions on the bottom hadronisation.
\end{enumerate}

This non-exhaustive list is just an excerpt of the potential of the experimental LHC heavy-ion program in the heavy-flavour sector connecting different measurements.  There is certainly the power to better constrain experimentally our picture of heavy-quark production as well as their interactions with deconfined matter. However, the full exploration of the data will require close collaboration between experiments and phenomenology in order to finally falsify part of the models and progressing towards a standard picture of heavy-quarks in nuclear collisions.

\section{Conclusions}
Future measurements promise to present qualitative as well quantitative improvements of the understanding of quarkonium modification by deconfinement. In addition, tests of perturbative QCD will be available in nuclear collisions in particular at low Bjorken-$x$. The full exploitation of the data will require good communication with the phenomenology and theory side in order to put the results in the context of other measurements and to benefit from uncertainty cancellations.
The establishment of a standard picture of quarkonia and heavy quark production and medium interaction in heavy-ion collisions at very high collision energies will be a difficult but rewarding task in the quest for the quark-gluon plasma.

\section*{Acknowledgements}

 The corresponding author is indebted for informative discussions on the subject with Anton Andronic, Ionut Arsene, Johanna Stachel, Torsten Dahms, Enrico Scomparin, Raju Venugopalan, Francois Arleo and Ulrich Uwer. Furthermore, in the direct preparation of the talk and these proceedings, the author thanks for prompt help concerning concrete numbers and references for the different experiments for available as well as upcoming data takings by Ionut Arsene, Enrico Scomparin, Roberta Arnaldi, Ginez Martinez, Giulia Manca, Patrick Robbe, Fr\'ed\'eric Fleuret, Camelia Mironov, Federico Antinori and Felix Reidt.
The corresponding author acknowledges support from the Studienstiftung des deutschen Volkes as a PhD fellow (until 31st of March) and from the European Research Council (ERC) through the project EXPLORINGMATTER, founded by the ERC through a ERC-Consolidator-Grant (since 1st of June).

\end{document}